\documentstyle[aps]{revtex}
\tighten
\begin{document}
\draft
\preprint{}
\title{On adiabatic turn-on and the asymptotic limit in linear response theory
for open systems}
\author{Jens U.~N{\"o}ckel and A.~Douglas Stone}
\address{
Applied Physics, Yale University, P.O. Box 2157,
Yale Station, New Haven CT 06520-2157}
\author{Harold U.~Baranger}
\address{
AT\&T Bell Laboratories (1D-230)\\
600 Mountain Ave., Murray Hill, New Jersey 07974-0636}
\date{}
\maketitle
\begin{abstract}

Linear response theory for open (infinite) systems leads to an expression
for the current response which contains surface terms in addition to
the usual bulk Kubo term.  We show that
this surface term vanishes identically if the correct order of limits
is maintained in the derivation: the system size, L, must be taken to infinity
before the adiabatic turn-on rate $\delta$ of the perturbation is taken to
zero.
In contrast to recent claims this shows that linear response theory for
open systems is gauge-invariant without modification of the continuity
equation.  We show that a simpler derivation of the Landauer-B\"uttiker
equations may be obtained consistently from the bulk Kubo term, noting that
surface terms arising here are non-vanishing because they involve the
opposite order of limits, $\delta \to 0$, then $L \to \infty$.

\end{abstract}
\pacs{72.10.Bg, 05.60.+w}

The connection between resistance and scattering properties of an open
phase-coherent system, pioneered by Landauer \cite{landauer} and
B\"uttiker \cite{butti}, has proved to be
an invaluable tool in understanding the wealth of new transport phenomena
that are observed in mesoscopic systems \cite{been}.
Given the $S$ matrix elements for scattering from subband $a$ in lead $n$ to
subband $a'$ in lead $m$, one can relate the currents and voltages in the leads
by $I_m=\sum_n g_{mn}\,V_n$,
where the conductance coefficients are determined for $m\ne n$ by
\begin{equation}\label{land}
g_{mn}=\frac{e^2}{h}\int\!\!dE\,\,
\left[-f'(E)\right]\sum_{a'a}\left|S_{mn,a'a}\right|^2
\end{equation}
($g_{mm}$ follows from current conservation). The original derivation of this
multiprobe formula by B{\"u}ttiker \cite{butti}
for $T=0$ is independent of the presence of
a homogeneous magnetic field, because the cancellation between group velocity
and density of states on which it relies still occurs at $B\ne 0$.
The simplicity and generality of Eq.\ (\ref{land}) made it desirable to
establish whether it followed from standard linear response theory (LRT)
applied
to open systems \cite{Sto88}.
For the case of multilead structures in a
non-zero magnetic field
this was first shown by Baranger and Stone \cite{baranger}, and subsequently
several different treatments were given \cite{shepard,janssen}.

Shortly thereafter, Sols \cite{sols} challenged the validity of
these derivations, arguing that due to a boundary term arising only
for {\em open systems}, the Kubo formula could not be derived
consistently without inserting
an additional term into the continuity equation.  Indeed it was argued
that this surface term violated gauge-invariance (since it vanished
in some gauges and supposedly did not in others) and that the more general
requirement of gauge-invariance necessitates the proposed modification of
the continuity equation \cite{sols}. Actually
Appendix D of Ref. \cite{baranger}
presented a proof that the surface term in question vanished identically
in the non-interacting limit.
The inconsistency according to Sols arose because subsequent
manipulations leading to the B\"uttiker equations give rise to surface
terms which were found to be non-zero and yet apparently had the same
form as those shown to vanish in Appendix D.
We have reexamined this issue and the closely related question of the
invariance of LRT under global gauge transformations.
Our results resolve the apparent contradiction by clarifying
that the two limits which arise for LRT in open systems, the limit of
the infinite system size and the adiabatic turn-on of the
perturbation {\em do not commute}.  The two apparently identical surface
terms noted by Sols differ precisely in the interchange of these limits
and we show that they vanish in one order and not in the other as found
in Ref. \cite{baranger}.  Hence Eq.\ (\ref{land}) follows rigorously from LRT;
a minor modification of our derivation proves the
gauge-invariance of LRT without modification of the continuity
equation.  In addition our analysis leads to a derivation of
Eq. (\ref{land}) which is somewhat simpler than that in
Ref. \cite{baranger} since
it only employs properties of the scattering wavefunctions and makes clear
the role of the adiabatic turn-on in breaking time-reversal symmetry.
This derivation is similar to that given in Ref.\ \cite{shepard} except that
we identify the central role of the above limits, clarifying in this way
both the conceptual and technical steps leading to Eq. (\ref{land}).

In Refs.\ \cite{Sto88,baranger,shepard,janssen,sols}, an open conductor is
represented by a finite sample region connected to $N$ straight
semi-infinite leads on which the external potential
$\phi$ assumes constant but
unequal values. We start with an adiabatically switched-on
DC perturbation of the form
\begin{equation}\label{perturb}
V(t)=e\int\rho({\bf r})\,\phi({\bf r})\,e^{\delta t}\,d{\bf r}.
\end{equation}
where $\rho({\bf r})$ is the electron density operator.
The expectation value of the current density
caused by the perturbation in linear order can be written as
\cite{sols}
%\widetext
\begin{equation}\label{linresp0}
{\bf J}({\bf r})=-e\lim_{\delta\to 0}\int\limits_{-\infty}^0
\!\!\!dt'e^{\delta t'}\int\limits_{0}^{\beta}\!\!d\lambda\,\int\!\!\!d{\bf r}'
\left\langle\,
{\bf j}({\bf r},-i\hbar\lambda)\,\dot{\rho}({\bf r}',t')
\right\rangle\,\phi({\bf r}').
\end{equation}
%\narrowtext
Applying the continuity equation
$e\dot{\rho}({\bf r},t')=-\nabla{\bf j}({\bf r},t')$ in the interaction picture
and integrating by parts, one arrives at
%\widetext
\begin{equation}\label{linresp}
{\bf J}({\bf r})=\lim_{\delta\to 0}\int\limits_{-\infty}^0
\!\!\!dt'e^{\delta t'}\int\limits_{0}^{\beta}\!\!d\lambda\,\left\langle\,
{\bf j}({\bf r},-i\hbar\lambda)\left[-\int\!\!\!d{\bf r}'
{\bf j}({\bf r}',t')\cdot{\bf\nabla}'
\phi({\bf r}')+\int\limits_{s'\to\infty}
\!\!\!d{s}'\, j_{\perp}({\bf s}',t')\,\phi({\bf s}')
\right]\right\rangle,
\end{equation}
%\narrowtext
where $j_{\perp}$ is the normal component of the current density, and
the surface ${\bf s}'$ must be taken to infinity because the original
integral, Eq.\ (\ref{linresp0}), is of infinite range for the open system.
However in the bulk term, which we denote by ${\bf J}^{K}({\bf r})$ because it
leads to the Kubo formula, the
spatial integral only has to be extended over the
{\em finite} region $\cal A$ where the perturbing electric field
${\bf E}=-{\bf\nabla}'\phi$ is nonzero.
To bring out the issue very clearly,
let us now integrate by parts in the bulk term ${\bf J}^{K}({\bf r})$,
noting that $\cal A$ is finite so the surface $\delta {\cal A}$ is not
to be taken to infinity.  We obtain
%\widetext
\begin{equation}\label{kubocont}
{\bf J}^{K}({\bf r})=\lim_{\delta\to 0}\int\limits_{-\infty}^0\!\!\!dt'
e^{\delta t'}\int\limits_{0}^{\beta}\!\!d\lambda
\,\left\langle\,
{\bf j}({\bf r},-i\hbar\lambda)\left[-e\int\limits_{\cal A}\!\!\!d{\bf r}'
\dot{\rho}({\bf r}',t')\phi({\bf r}',t')-\int\limits_{\partial{\cal A}}
\!\!\!d{s}'\,j_{\perp}({\bf s}',t')\,\phi({\bf s}',t')
\right]\right\rangle.
\end{equation}
%\narrowtext
The surface term in this equation differs from the one in Eq.\ (\ref{linresp})
only in sign and in the absence of the $s'\to\infty$ limit. Now we can
determine the total current $I_m$
in lead $m$ by integrating ${\bf J}({\bf r})$ over a cross section of lead $m$.
Let $I^K_m$and $I_m'$ denote the contributions of
${\bf J}^K$ and the surface term in Eq.\ (\ref{linresp}) to $I_m$.
The subtle point emerges once we take the limit $\delta \to 0$ in the above
expression for $I^K_m$, as
the property $\nabla\cdot{\bf J}^K=0$ and the nonlocal Onsager relation for
$\sigma_{DC}$ in a magnetic field imply that the bulk term in
Eq.\ (\ref{kubocont}) vanishes upon integration over a lead cross section,
leaving only the surface term \cite{baranger}.
This is physically reasonable because
the DC currents in the leads are uniquely determined by the
voltages outside of $\cal A$.
But since the only restriction on $\cal A$ was that ${\bf E}=0$ outside of it,
nothing now
prevents us from taking $\partial{\cal A}\to\infty$. Thus $I^K_m$
becomes identical to $-I'_m$ {\em except for the
interchange of the limits}.  If these two limits commuted we would
obtain the absurd result $I_m=I^K_m+I'_m=0$ !

Note first that the bulk term in $I^K_m$
only vanishes when we take the
limit $\delta\to 0$, so the order of limits leading to the surface
expression for $I^K_m$ is uniquely determined. Conversely,
the limit $s'\to\infty$ in
Eq.\ (\ref{linresp}) must be taken {\em before} $\delta$ can go to zero. The
reason is that in calculating the LRT current as a trace over the unperturbed
eigenstates, we assume that the system was in equilibrium in the distant past,
which is achieved by $\delta>0$. Matrix elements of the perturbation,
Eq.\ (\ref{perturb}), in the unperturbed basis give rise to
the spatial integral of infinite range. This is the asymptotic limit, which
must consequently precede the adiabatic limit, $\delta\to 0$.
It is not immediately obvious that this
subtle difference really matters; therefore, we now evaluate both surface terms
carefully, to show that $I_m'$ vanishes while $I_m^K$ does not.
Inserting many-body eigenstates
$\vert\alpha\!\!>$ and executing the integrals over $t'$ and $\lambda$,
one obtains from Eq.\ (\ref{kubocont})
%\widetext
\begin{equation}\label{linresp1}
I_m^K=\int\!\!dy_m\,{\bf J}^K({\bf r})=i\hbar\int\!\!dy_m\,
\lim\limits_{\delta\to 0}
\sum\limits_{\alpha,\beta}
\frac{P_{\beta}-P_{\alpha}}{E_{\beta}-E_{\alpha}}
\frac{1}{E_{\beta}-E_{\alpha}+i\hbar\delta}
\int\limits_{\partial{\cal A}}\!\!\!d{s}'
<\!\!\beta\vert\,{\bf j}({\bf r})\vert\alpha\!\!>\,<\!\!\alpha\vert
j_{\perp}({\bf s}')\vert\beta\!\!>\,\phi({\bf s}'),
\end{equation}
%\narrowtext
where the $y_m$ integral extends over a cross section of lead $m$.
The expression for $I'$ is the same,
only with $\partial{\cal A}$ replaced by $s'\to\infty$.

In the non-interacting limit, the statistical weights $P_{\alpha}$ can be
replaced by Fermi functions $f(E_{\alpha})$, and $\vert\alpha\!\!>$ become
single-particle scattering states which, in lead $l$, have the asymptotic form
%\widetext
\begin{equation}\label{scatt}
\psi_{\alpha}^l\equiv\psi_{Ea{p}}^l\,\equiv\,\delta_{{p} l}
\,\xi_{-a}^l+\sum_{a'}S_{l{p},a'a}\,\xi_{+a'}^l.
\end{equation}
%\narrowtext
Here, the label $\alpha$ consists of the energy $E$, subband index $a$ and lead
$p$ of the incident wave. An analogous definition holds for
$\psi_{\beta}^l\equiv\psi_{E'b{q}}^l$.
The quantum wire eigenfunctions have the form
\begin{equation}\label{efn}
\xi_{\pm a}^l({\bf r})= \left|2\pi\,\frac{dE_a^l}{dk}\right|^{-1/2}
e^{i{\rm k}_{\pm a}^lx}\,\chi_{\pm a}^l(y),
\end{equation}
where $\chi_{\pm a}^l$ are the transverse wavefunctions and
${\rm k}_{\pm a}^l$ is the outgoing ($+$) or incoming ($-$) wavenumber. With
the above choice of normalization,
the symbolic sums over collective labels have the explicit form
$\sum_{\alpha}\to\sum_p\sum_a\int\!\!dE$.
In the following, subscripts $a$ imply a dependence on the energy
variable $E$, while indices $b$ belong to energies $E'$.
As was shown in Ref.\cite{baranger}, current conservation
implies the following properties for two wire eigenfunctions at
the same energy in the asymptotic region:\cite{factor}
\begin{equation}
\int\!\!dy_l\,
(\xi_{\sigma a}^{l}\vert j_{\perp}({\bf r})\vert\xi_{\sigma'a'}^{l})
=\sigma \frac{e}{2\pi\hbar}\,\delta_{\sigma'\sigma}\delta_{a'a}
\label{currcons1}
\end{equation}
with the notation $\sigma=\pm$, and round brackets for the single particle
current matrix elements. If $\epsilon_a^{{l}}$
is the propagation threshold of subband $a$ in lead $l$,
Eq.\ (\ref{linresp1}) leads to integrals of the type
\begin{equation}\label{ff}
\int\limits_{\epsilon_a^{{p}}}^{\infty}\!\!dE\,\,
\int\limits_{\epsilon_b^{{q}}}^{\infty}\!\!dE'\,\,
\frac{f(E')-f(E)}{E'-E}
\,\frac{F_S(E',E)}{E'-E+i\hbar\delta}
\end{equation}
where
\begin{equation}\label{sixteen}
F_S(E',E)\equiv
\int\limits_{S}\!\!\!d{s}'
(\psi_{E'b{q}}^m\vert j_{\perp}({\bf r})\vert\psi_{Ea{p}}^m)
(\psi_{Ea{p}}^n\vert j_{\perp}({\bf s}')\vert\psi_{E'b{q}}^n)
\end{equation}
contains the current matrix elements and thus vanishes at threshold. The
surface $S$ of integration is $\partial{\cal A}$ for $I_m^K$, and is taken to
infinity for $I'$. Although we wish to let $\partial{\cal A}\to\infty$
eventually, we can do so only after $\delta\to 0$.
Note also that $\frac{f(E')-f(E)}{E'-E}$ is non-singular on the real axis.

We now evaluate the integral over $E'$ in Eq.\ (\ref{ff}) by Cauchy's theorem.
In general the integrand in Eq.\ (\ref{ff}) may have a complicated singularity
structure in the complex plane, arising both from the ``Matsubara'' poles of
the first factor and the singularities of the S-matrix contained in
$F_S(E,E')$.
However since asymptotically the wavefunctions have a plane-wave dependence
on $x'$ (the longitudinal coordinate in the lead), we shall see that
contributions from the return contour may always be made to vanish as long
as the correct half-plane is chosen to close the contour. This is so even
if the return contour is a finite (not infinite) distance from the real
axis \cite{shepard2}.
Since the locations of all the singularities of the integrand in
Eq. (\ref{ff})
are independent of both the asymptotic and adiabatic limits except
for the pole at $E'=E-i\hbar\delta$, for convenience we choose contours
$C_1$, $C_2$ (shown schematically in Fig. 1) which by assumption enclose only
the latter pole.
Hence $C_1$ will contain no singularities at all,
and terms which in the asymptotic limit may be closed in the upper-half
plane will give zero contribution.   $C_2$ will enclose the pole at
$E'=E-i\hbar\delta$ unless it approaches the real axis outside the range
of the $E'$ integration (i.e. unless $E$ is less than the sub-band threshold
relevant for the term in question).  By this convenient choice of contours
we need only evaluate the residue at $E'=E-i\hbar\delta$ and see how it
depends explicitly on the order of the asymptotic vs. adiabatic limits.

The residue theorem yields with $\nu=1,2$
%\widetext
\begin{eqnarray}
&&\left(\int\limits_{\epsilon_b^{{q}}}^{\infty}\!\!dE'\,\,+
\int\limits_{C_{\nu}}\!\!dE'\,\,\right)
\,\frac{f(E')-f(E)}{E'-E}
\,\frac{F_S(E',E)}
{E'-E+i\hbar\delta}\label{contint}=\\
&&-2\pi i\,\delta_{\nu,2}\,\Theta(E-\epsilon_b^q)\,
\frac{f(E-i\hbar\delta)-f(E)}{-i\hbar\delta}
\,F_S(E-i\hbar\delta,E)\nonumber,
\end{eqnarray}
%\narrowtext
where the step function $\Theta$ enters since
$C_2$ encloses a pole only if $E>\epsilon_b^q$, and
$\delta_{\nu,2}$ reflects the fact that
no poles are enclosed by $C_1$.
The additional minus sign is due to the negative sense in which $C_2$ encloses
the pole. Along $C_{\nu}$, the limit $S\to\infty$ enables us to apply
the asymptotic forms of $\psi_{\alpha}$ and $\psi_{\beta}$
in the second current matrix element in $F_S(E',E)$ to get
in lead $l$
\begin{eqnarray}\label{prod}
(\psi_{Ea{p}}^l\vert
j_{\perp}({\bf s}')\vert\psi_{E'b{q}}^l)
&=&\delta_{{q} l}\,\delta_{{p} l}\,(\xi_{-a}^{l}\vert
j_{\perp}\vert\xi_{-b}^{l})+
\delta_{{p} l}\sum\limits_{b'}S_{l{q},b'b}(\xi_{-a}^{l}\vert
j_{\perp}\vert\xi_{+b'}^{l})\\
&&+\delta_{{q} l}\sum\limits_{a'}S_{l{p},a'a}^{*}
(\xi_{+a'}^{l}\vert
j_{\perp}\vert\xi_{-b}^{l})+
\sum\limits_{a'b'}S_{l{p},a'a}^*S_{l{q},b'b}
(\xi_{+a'}^{l}\vert
j_{\perp}\vert\xi_{+b'}^{l}),\nonumber
\end{eqnarray}
{\em irrespective} of whether $\delta=0$ or not in Eq.\ (\ref{contint}).
For $E'$ in the complex plane, the imaginary parts of the wavenumbers satisfy
\begin{equation}\label{sign}
{\rm sgn}\left({\rm Im} \,{\rm k}_{\pm b}^l(E')\right)=
\pm {\rm sgn}\left({\rm Im}\,E'\right).
\end{equation}
This can easily be checked for the special case
${\rm k}_{\pm b}^l\propto\pm\sqrt{E'-\epsilon_b^{q}}$, but holds even in
asymmetric leads at $B\ne 0$ where $E'=\epsilon_b^{q}$ can occur for nonzero
$k$. Note that the branch point at $\epsilon_b^{q}$ causes no problems since
$F_S(E',E)$ vanishes there. The first and third term of Eq.\ (\ref{prod})
depend on $\xi_{-b}^{l}\propto e^{i{\rm k}_{-b}^lx'}$, so that they acquire an
exponential decay factor $e^{-\kappa(E') x'}$ when integrated along
the contour $C_2$. For the other two terms in Eq.\ (\ref{prod}),
we choose $C_1$ to obtain positive imaginary parts in
${\rm k}_{+b}^l(E')$. The asymptotic limit
$x'\to\infty$ causes the contributions from $C_{1}$ and $C_{2}$
to vanish identically. The result is that the real
axis integrals over $E'$ in Eq.\ (\ref{ff})
vanish due to Cauchy's theorem unless $C_2$ encloses a pole. Its residue in
Eq.\ (\ref{contint}) contains a factor
$e^{i{\rm k}_{-b}^l(E-i\hbar\delta)x'}$. For $E\ne\epsilon_b^q$ we can
expand the outgoing wavenumber as ${\rm k}_{-b}^l(E)-i\delta\,v^{-1}(E)$
where the group velocity $v(E)=\frac{1}{\hbar}\frac{dE}{dk}$ is
negative. We thus pick up a decaying
exponential of the form
\begin{equation}\label{noncommute}
F_S(E-i\hbar\delta,E)\propto \exp\left[{-\delta\,\left|v^{-1}(E)\right|\,x'}
\right].
\end{equation}
This exhibits clearly the
noncommutability of the limiting procedures: In $I_m'$ we take $x'\to\infty$
first, causing Eq.\ (\ref{noncommute}) and with it $I_m'$ itself to vanish.
Moreover, since all boundary terms encountered in a gauge transformation
\cite{sols}
contain Eq.\ (\ref{ff}) with the above order of limits, this observation
establishes the
gauge invariance of LRT.

On the other hand, $I_m^K$ is nonzero because now $\delta\to 0$ is
performed first in Eq.\ (\ref{noncommute}), yielding
%\widetext
\begin{eqnarray}
&&\lim_{\delta\to 0}
\int\limits_{\epsilon_b^{{q}}}^{\infty}\!\!dE'\,\,
\,\frac{f(E')-f(E)}{E'-E}
\,\frac{F_S(E',E)}
{E'-E+i\hbar\delta}\label{contint1}\nonumber\\
&=&-2\pi i\,\delta_{\nu,2}\,\Theta(E-\epsilon_b^q)\,f'(E)\,F_S(E,E)
\label{residue}
\end{eqnarray}
%\narrowtext
This result leads to Eq.\ (\ref{land}), as we now show by
taking the asymptotic limit $S\to\infty$.
The integral over ${s}'$ in Eq.\ (\ref{linresp1})
becomes a sum of integrals over isolated quantum
wire cross sections, $\sum_n\int\!\!dy_n$, along which $\phi$ assumes
its constant values $V_n$.
Our expression for the total current at this stage already has the form
$I_m=I_m^K=\sum_n g_{mn}\,V_n$.
The product of current matrix elements at the same energy in $F_S(E,E)$
can be evaluated with Eq.\ (\ref{prod}) if we let $\bf r$
in Eq.\ (\ref{linresp1}) go to infinity as well.
Considering only $n\ne m$, 7 of the 16 terms in $F_S(E,E)$
vanish identically due to Kronecker delta factors like
$\delta_{pm}\delta_{pn}$. When the sums over
the incident wave parameters $p,a$ or $q,b$ are performed as prescribed by
Eq.\ (\ref{linresp1}), five more terms drop out due to the unitarity
of the $S$ matrix,
\begin{equation}\label{unitarity}
\sum_{{p}=1}^N {\sum_c}^E
S_{n{p},a'c}^*S_{m{p},ac}=\delta_{nm}\delta_{a'a}=0.
\end{equation}
Here, we sum over all subbands propagating at energy $E$.
This leaves the following four terms in $F_S(E,E)$:
%\begin{mathletters}
\begin{eqnarray}
&&\delta_{{q} m}\delta_{{p} m}S_{nm,a'a}^*S_{nm,b'b}
\left(\xi_{-b}^{m}\vert j_{\perp}\vert\xi_{-a}^m\right)
\left(\xi_{+a'}^{n}\vert j_{\perp}\vert\xi_{+b'}^n\right)+\nonumber\\
&&\delta_{{q} m}\delta_{{p} n}S_{mn,a'a}S_{nm,b'b}
\left(\xi_{-b}^{m}\vert j_{\perp}\vert\xi_{+a'}^m\right)
\left(\xi_{-a}^{n}\vert j_{\perp}\vert\xi_{+b'}^n\right)+
\nonumber\\
&&\delta_{{p} m}\delta_{{q} n}S_{nm,a'a}^*S_{mn,b'b}^*
\left(\xi_{+a'}^{n}\vert j_{\perp}\vert\xi_{-b}^n\right)
\left(\xi_{+b'}^{m}\vert j_{\perp}\vert\xi_{-a}^m\right)+
\nonumber\\
&&\delta_{{q} n}\delta_{{p} n}S_{mn,b'b}^*S_{mn,a'a}
\left(\xi_{+b'}^{m}\vert j_{\perp}\vert\xi_{+a'}^m\right)
\left(\xi_{-a}^{n}\vert j_{\perp}\vert\xi_{-b}^n\right),\label{terms}
\end{eqnarray}
where a sum over propagating subbands $a'$, $b'$ must be performed.
The first two terms require integration along $C_1$ in Eq.\ (\ref{contint1})
and hence yield no residue.
For the third term, Eq.\ (\ref{currcons1}) yields zero.
In Ref.\ \cite{shepard}, it is not noticed that
this term does survive the contour integration and only vanishes
due to Eq.\ (\ref{currcons1}), which is not equivalent to the unitarity
of the $S$ matrix, but follows instead from the translational invariance of the
asymptotic wire Hamiltonians. That this property of the model is important in
deriving Eq.\ (\ref{land}) was recognized solely in Ref.\ \cite{baranger}.
In the present treatment, the remarks below Eq.\ (\ref{sign}) allow us to
conclude that no other symmetries of the leads are essential.

Our expression for $I_m^K$ now contains only the
last term in Eq.\ (\ref{terms}). The conductance coefficients for
$m\ne n$ are then given by
%\widetext
\begin{eqnarray}\label{land1}
g_{mn}&=&i\hbar\int\!\!dy_m\int\!\!dy_n
\sum_{{p,q}=1}^N \sum\limits_{a,b=1}^{\infty}(-2\pi i)
\int\limits_{\epsilon_a^{{p}}}^{\infty}\!\!dE\,
\Theta(E-\epsilon_b^q)\,f'(E)\\
&&\times\delta_{{q} n}\delta_{{p} n}\sum\limits_{a'b'}
\left[S_{mn,b'b}^*S_{mn,a'a}
\left(\xi_{+b'}^{m}\vert j_{\perp}\vert\xi_{+a'}^m\right)
\left(\xi_{-a}^{n}\vert j_{\perp}\vert\xi_{-b}^n\right)\right]_{E'=E}
\nonumber\\
%&=&2\pi\hbar
%\int\limits_{\epsilon_1^{{n}}}^{\infty}\!\!dE\,\,f'(E)
%{\sum\limits_{a'ab'b}}^{E}S_{mn,b'b}^*S_{mn,a'a}
%\left(\frac{e}{2\pi\hbar}\right)\left(-\frac{e}{2\pi\hbar}\right)
%\delta_{b'a'}\delta_{ab}\nonumber\\
&=&\frac{e^2}{h}\int\limits_{\epsilon_1^{{n}}}^{\infty}\!\!dE\,\,\left[-f'(E)
\right]{\sum_{a'a}}^{E}\left|S_{mn,a'a}\right|^2,\nonumber
\end{eqnarray}
%\narrowtext
where Eq.\ (\ref{currcons1}) has been used.
This is the desired result, Eq.\ (\ref{land}).
Comparing the first and last term in Eq.\ (\ref{terms}),
we see that they differ only in the order of $m$ and $n$.
It is the sign of $\delta$ that caused the first term to make no contribution
to Eq.\ (\ref{residue}), so that $g_{mn}$
depends only on scattering probabilities from $n$ to $m$, and not
{\em vice versa}, as required by causality. Finally we comment that the
vanishing of $I_m'$ can also be demonstrated in the presence of
interactions: the
essential point is the smoothness of the integrand in Eq.\ (\ref{ff}) on the
real axis for $\delta\ne 0$, which is preserved in the interacting case.

This work was supported under NSF grant DMR 9215065.

\narrowtext
\begin{figure}
\caption{
\label{branch}The two possible contours that yield exponential decay in
$x'$ off the real axis for the various terms in Eq.\ (\protect\ref{prod}).
Open dots: two possible locations of the
pole at $E'=E-i\hbar\delta$}
\end{figure}


\begin{references}
\bibitem{landauer}
R.~Landauer, Z.~Phys. {\bf 68}, 217 (1987)
and references therein
\bibitem{butti}
M.~B{\"u}ttiker, Phys.~Rev.~Lett. {\bf 57}, 1761 (1986);
M.~B{\"u}ttiker, IBM J.~Res.~Develop. {\bf 32}, 317 (1988)
\bibitem{been}For a review see
C. W. J. Beenakker and H. van Houten in
{\it Solid State Physics}, Vol. 44, edited by H. Ehrenreich and D. Turnbull
(Academic Press, New York, 1991) pp. 1-228
\bibitem{Sto88}
A. D. Stone and A. Szafer, IBM J. Res. Develop. {\bf 32}, 384 (1988)
and references therein
\bibitem{baranger}
H.~U.~Baranger and A.~D.~Stone, Phys.~Rev.~B {\bf 40}, 8169 (1989)
and references therein
\bibitem{shepard}
K.~Shepard, Phys.~Rev.~B {\bf 43}, 11623 (1991)
\bibitem{janssen}
M.~Jan{\ss}en, Solid State Comm. {\bf 79}, 1073 (1991)
\bibitem{sols}
F.~Sols, Phys.~Rev.~Lett. {\bf 67}, 2874 (1991)
%\bibitem{taylor}
%J.~R.~Taylor, {\em Scattering Theory} (John Wiley, New York, 1972)
\bibitem{factor}
The factor in Eq.\ (\ref{currcons1}) differs from Ref.\ \cite{baranger} by
$1/h$ because the normalization in Eq.\ (\ref{efn}) lacks a factor of
$\sqrt{h}$.
\bibitem{shepard2}
Ref.\ \cite{shepard} creates the impression that the return
contour simply has to be taken as an arbitrarily large semicircle which
encloses no poles. We point out that such a
contour encloses the poles of $f(E')$ and
extends to $-\infty$ on the real axis, all of which cannot be justified
without the asymptotic limit. In addition,
Eq.\ (24) for the conductance coefficients in that paper has the wrong sign.
That mistake is
canceled by a sign error in one of the current matrix elements.
\end{references}
\end{document}